 \documentclass[final,3p,times,twocolumn]{elsarticle}

\usepackage{amssymb}

\begin{document}

\begin{frontmatter}



\title{Heterogeneity and Disorder: Contributions of Rolf Landauer}


\author[Harvard]{Bertrand I. Halperin}
\author[TAU]{David J. Bergman}

\address[Harvard]{Department of Physics, Harvard University, Cambridge, MA
02138, USA}

\address[TAU]{Raymond and Beverly Sackler School of Physics \& Astronomy,
Tel Aviv University, IL-69978 Tel Aviv, Israel}

\date{\today}


\begin{abstract}
Rolf Landauer made important contributions to many branches of science.  Within the broad area of transport in disordered media, he wrote seminal papers on electrical conduction in macroscopically inhomogeneous materials, as well as fundamental analyses of electron transport in quantum mechanical systems with disorder on the atomic scale. We review here some of these contributions. 

\end{abstract}


\end{frontmatter}



\section{Introduction}
\label{introduction}

Rolf Landauer was recognized for outstanding accomplishments in
many branches of science.  In addition to his work on transport in
inhomogeneous systems, which will be the focus of the present article,
Landauer wrote papers on noise and fluctuations, on nonlinear wave
propagation and soliton formation, on ferroelectric instabilities and
displacive soft modes, entropy production in systems out of equilibrium,
philosophical principles of science and technology, and above all,
on the physical limits to computation.  His influence in this last
area was of such a magnitude that he  was the subject of a  ``Profile''
article by Gary Stix in the September 1998 issue of {\em Scientific
American} \cite{Stix}.
The article was titled ``Riding the Back of Electrons'', and
subtitled ``Theoretician Rolf Landauer remains a defining figure in the
physics of information.''

During his lifetime, Landauer received many awards for his work including
the Ballantine Medal of the Franklin Institute in 1992, the  1995 Buckley
Prize of the American Physical Society,  the LVMH, Inc.\ Science for Art
Prize in 1997, the 1998 IEEE Edison Medal, an honorary doctorate from the
Technion in 1991, and a Centennial Medal from Harvard University in 1993.
He was elected to the National Academy of Science, the National Academy
of Engineering, and the American Academy of Arts and Sciences in the US,
and to the European Academy of Arts.

\begin{figure}
\includegraphics[width=\columnwidth]{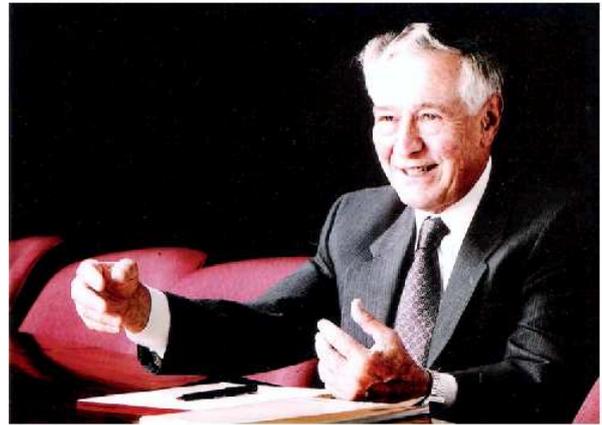}
\caption{Photograph of Rolf Landauer. Printed with permission from
his survivors.
\label{LandauerPhoto}}
\end{figure}

Landauer's influence on science and technology was not limited to the
importance of his scientific discoveries and research.  He had much
to say about the conduct of research and about philosophical issues
of how science should be interpreted.  He also delighted in challenging
entrenched ideas and in forcing people to think more carefully   about the
foundations of their work---see Fig.\ \ref{LandauerPhoto}
for a typical appearance of Rolf Landauer in this mode.
There is perhaps no better way to illustrate
this aspect of his character than to cite some of the titles of articles
that Landauer wrote in the last decade of his life: ``Light faster than
light'' \cite{L157}, ``Is quantum mechanics useful?'' \cite{L161},
``Mesoscopic noise: Common sense view'' \cite{L171},
 ``Zig-zag  path to understanding'' \cite{L164}, ``Conductance is
 transmission'' \cite{L174},
``The physical nature of information'' \cite{L173}, and ``Fashions in
science and technology'' \cite{L175}.

Due to limitations of space, our  discussion of  Landauer's scientific
contributions will be restricted  to his work on transport in inhomogeneous
systems, and some closely related work on quantum mechanical effects
in mesoscopic systems. However, we include at the end of the article a
brief summary of Landauer's personal history.

\section{Landauer and inhomogeneous systems}

Although major advances in the understanding of electrical conductivity
of disordered and heterogeneous media were made by a number of
his contemporaries, Landauer's approach was unique in this field:
On one hand, people like William Fuller Brown, Jr.\
\cite{WFBrown55} or Zvi Hashin and
Shmuel Shtrikman \cite{HashinShtrikmanJAP62}
confined themselves to a discussion of systems
where a classical physics approach is valid, and described the
local electrical response in terms of a position dependent conductivity.
This lead, eventually, to concepts like percolation threshold, which
determines the macroscopic response of a metal/insulator mixture
\cite{AharonyStaufferPercolation}.
On the other hand, people like Philip W. Anderson \cite{PhilAndersonPRB68}
and Neville F. Mott \cite{MottAdvPhys67,HoriBook68} focussed
upon the effects of microscopic disorder on the detailed form of the
quantum mechanical wave function. This lead, eventually, to concepts like
Anderson localization and Mott transition as governing the macroscopic
response of such a system. By contrast, Landauer often tried to
combine quantum considerations with classical physics considerations.
This is clearly evident in his work on electromigration
\cite{LandauerMigration57,ResistivityDipole75},
but also in his classic paper which derived the famous ``Landauer
Formula''
\cite{LandauerFormulaPhilMag69}, where he used the classical physics
Einstein relation in order to derive the macroscopic conductivity of
a one dimensional disordered system from the diffusion coefficient of a
single electron. A similar combination of quantum and classical
approaches can also be found in Landauer's work on conductivity of
cold-worked metals \cite{LandauerColdWorkPR51} and on Lorentz
corrections to electrical conductivity \cite{LandauerLorentzPRB72}.

\section{Classical inhomogeneous systems}
\label{macro}

Rolf Landauer became interested in inhomogeneous systems early on in
his career. In 1952
he published a paper entitled ``The electrical resistance of binary
metallic
mixtures'' \cite{SEMALandauer}, where he developed a simple approximation
for calculating that resistance, which is a macroscopic property of such
mixtures. This approximation yields the following equation for   the
macroscopic scalar conductivity $\sigma_{\rm eff}$ of a
multi-constituent, $d$-dimensional, isotropic composite medium in terms of
the constituent scalar conductivities
$\sigma_i$  and constituent volume fractions
$p_i$,  with $\sum_i p_i=1$:
\begin{equation}
0= \sum_i p_i \, \frac{\sigma_{\rm eff}-\sigma_i}{\sigma_{\rm
eff}+(d-1)\sigma_i}\;.
\end{equation}
In the case where there are just  two components, this becomes a quadratic
equation, which has the explicit solution
\begin{eqnarray}
\lefteqn{{2\, \frac{d-1}{d} \, \sigma_{\rm eff}=
 \sigma_1\left(p_1-\frac{1}{d}\right)+
  \sigma_2\left(p_2-\frac{1}{d}\right)}}\nonumber\\
&&+\,\left\{
 \left[\sigma_1\left(p_1-\frac{1}{d}\right)
 +\sigma_2\left(p_2-\frac{1}{d}\right)\right]^2\right.\nonumber\\
&&\;\;\;\;\left.+\;4\,\left.\frac{d-1}{d^2}
  \,\sigma_1\sigma_2\right.\right\}^{{1}/{2}}\;.
\label{SEMAcondutivity}
\end{eqnarray}


This approximation is still widely used, especially in the
context of a disordered
microstructure, and is generally known as the self-consistent (or
symmetric) effective medium approximation (SEMA). This approximation had
actually already been discovered
in 1935 by D. A. G. Bruggeman \cite{SEMABruggeman}, who was then a high
school teacher in the Netherlands. Landauer did not know about this, (nor
did the reviewer of his manuscript at J. Appl.\ Phys.) and he
therefore achieved this breakthrough independently. This approximation
can be contrasted with
an earlier approximation, known as the Clausius-Mossotti (CM) or Maxwell
Garnett approximation. In the latter approximation, $\sigma_{\rm eff}$
satisfies a linear algebraic equation which can be expressed in the
following concise form:
\begin{equation}
\frac{\sigma_{\rm eff}-\sigma_2}{\sigma_{\rm eff}+(d-1)\sigma_2}=
 p_1\frac{\sigma_1-\sigma_2}{\sigma_1+(d-1)\sigma_2}\, .
 \label{CM}
 \end{equation}
The CM result is non-symmetric in the two constituents: The $\sigma_2$
constituent plays the role of host while the $\sigma_1$ constituent
plays the role of inclusions. It is easy to generalize the CM result
to any number of different inclusion constituents that are embedded in one
common host constituent. This is achieved by rewriting $\sigma_2$ as
$\sigma_{\rm host}$, $\sigma_1$, $p_1$ as $\sigma_i$, $p_i$, and summing
the right hand side of the resulting equation over the different types of
inclusions $i$. This leads to
an equation that is still equivalent to a linear algebraic equation
for $\sigma_{\rm eff}$.

In contrast with CM, the SEMA result is
symmetric in all of the
constituents. When SEMA is extended to more than two constituents,
the result for $\sigma_{\rm eff}$ then becomes the solution of a
polynomial
equation with order equal to the number of constituents.

Interestingly, neither
Bruggeman nor Landauer realized, at first, that SEMA predicts
the existence of a conductivity threshold in the case where one of the
constituents is a perfect insulator. This crucial consequence, which does
not follow from the CM approximation, was first
appreciated by others \cite{PolderVanSanten}. Somewhat later it
was realized that the conductivity threshold in this system is
associated with a percolation threshold of the conducting
constituent \cite{LandauerUnpublishedNote},
which is a geometric property of the microstructure. This
threshold is a critical point, i.e., a singular point in
the physical response of the system as function of the physical parameters
\cite{AharonyStaufferPercolation}. This point is characterized by the
``percolation threshold'' $p_c=1/d$: When the volume fraction of the
conducting constituent $p_M$ is greater than $p_c$, the macroscopic
conductivity $\sigma_{\rm eff}$ is nonzero, but it vanishes when
$p_M\leq p_c$.
For $p_M\geq p_c$, $\sigma_{\rm eff}$ increases linearly with increasing
$p_M$, starting from 0:
\begin{equation}
\frac{\sigma_{\rm eff}}{\sigma_M}=
 \frac{d}{d-1}\left(p_M-\frac{1}{d}\right),\;\;\;
   p_M\geq p_c\equiv\frac{1}{d}\; ,
\label{LinearConductivity}
\end{equation}
where $\sigma_M$ is the scalar conductivity of the conducting constituent.
There is thus no discontinuity in the function
$\sigma_{\rm eff}(p_M)$, only a discontinuous slope at $p_M=p_c$. The
value of $p_c=1/d$ predicted by SEMA depends only on the dimensionality.
That and the linear
dependence of $\sigma_{\rm eff}$ upon $p_M$ are leading characteristics of
SEMA. In practice, both characteristics are rather inaccurate: Experiments
on real continuum composites 
show that the value
of $p_c$ depends on details of the microstructure \cite{HeaneyPRB95}.
Only in the case of a two-dimensional ($d=2$) disordered composite medium
where the microstructure is symmetric in the two constituents is the
value $p_c=0.5$, predicted by SEMA,
correct. The linear form of $\sigma_{\rm eff}(p_M)$,
predicted by SEMA for $p_M\geq p_c$, is also usually contradicted by
experiments on real composites and discrete network models, although the
SEMA prediction that $\sigma_{\rm eff}(p_M)$ is continuous at $p_c$ is
verified. In reality, the behavior of $\sigma_{\rm eff}(p_M)$ for small
positive values of $p_M-p_c$ is well described by a power law
$\sigma_{\rm eff}(p_M)\propto\sigma_M (p_M-p_c)^t$, where the ``critical
exponent'' $t$
has values that depend on general properties of the microstructure but
not on minute details---that is known as ``universality''. For example, in
discrete network models with finite-range-correlated randomness, it is
found that $t\approx 1.3$ when $d=2$, $t\approx 2.0$ when $d=3$,
$t=0.5$ when $d\geq 6$ \cite{BergStroudSSP92}. In continuum composites,
the value of $t$ also
sometimes depends on details of the microstructure, e.g.,
in the cases of ``swiss cheese'' and ``inverse swiss cheese'' models
of a conductor/insulator mixture \cite{HalperinFengSenPRL85}, and
in the case of a singular distribution of conductances in a random
resistor network \cite{KogutStraleyConductivityDistribution79}. In
any case,
in contrast with the SEMA prediction, $t$ is never equal to 1.

These failures of SEMA are related to the fact that it
is an uncontrolled approximation which cannot be improved in any
systematic
fashion: SEMA is based on a simple, intuitive
physical idea, namely, that when trying to calculate the electric
field and
current in and near a single spherical inclusion with conductivity
$\sigma_1$ or $\sigma_2$ one can replace the
rest of the heterogeneous system by a fictitious, uniform host with
$\sigma_{\rm eff}$ as its uniform conductivity. The value of this
initially
unknown macroscopic or ``bulk effective'' conductivity $\sigma_{\rm
eff}$ is
then found by imposing the self-consistency requirement that the
dipole current source, which is excited when an external uniform electric
field is imposed on any isolated spherical inclusion in this fictitious
uniform host, yields zero when summed over all the different inclusions
in the system. While this approximation becomes
exact when $\sigma_1\rightarrow\sigma_2$
or when the system is dilute, i.e., when either $p_1\ll 1$ or $p_2\ll 1$,
it is
impossible to estimate the error when neither of these conditions is
satisfied. The failure of SEMA in predicting correct values for
critical exponents, like $t$, and its inability to include relevant
details of the microstructure, as in the case of the above mentioned
swiss cheese model, have lead many scientists to abandon this
approximation. Instead, they chose to use techniques like
renormalization group transformation or brute force simulation of
discrete, random network models in order to study the macroscopic
response near a percolation threshold. (A review of the different
approaches to calculations of the macroscopic conductivity of a
composite medium, including cases where the system is near a
percolation threshold, can be found in Ref.\ \cite{BergStroudSSP92}.)
However, extension of SEMA to the case
where a strong magnetic field
is applied to a macroscopic mixture of two conductors with different
but comparable resistivities \cite{Stach}--\cite{SEMAcritPtPRB},
or to a mixture of three constituents where
one is a normal conductor while the other two are a perfect
insulator and a perfect conductor
\cite{BergStrelMISColumnarComposites}--\cite{BergmanPRB2001},
have resulted in the discovery of some new critical points which are
unrelated to the geometric percolation threshold. This demonstrates
a great advantage of SEMA and its extensions: Because they lead to closed
form expressions for the elements of the macroscopic resistivity tensor,
or at least to closed form (though complicated) coupled equations which
determine those elements, it is much easier
to identify a mathematical singularity in those moduli, which signals the
existence of a critical point.

\begin{figure}
\includegraphics[width=\columnwidth]{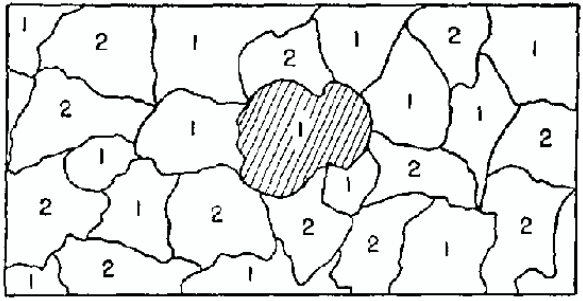}
\caption{Landauer's derivation of SEMA.  This figure from
 Ref.\ \protect\cite{SEMALandauer} was captioned,
``The shaded crystal of type
 1 is surrounded by crystals of both types, which are imagined to be
 replaced by a single medium of uniform conductivity." Landauer made
 the  further approximation of replacing the shaded crystal by a sphere.
Reprinted with permission from
The American Institute of Physics.
\label{LandauerEMA}}
\end{figure}

The research by Landauer described above was done while he worked
at the NACA Lewis
Laboratory, but by the
time Ref.\ \cite{SEMALandauer} appeared in print, he had moved to IBM.
Shortly afterwards,  Landauer got interested in the problem of
magneto-transport in a macroscopically heterogeneous or composite medium.
In a one page article, published in J. Appl.\ Phys.\ in 1956
\cite{LandauerHallJAP56} shortly after
the appearance of
a 1955 IBM Technical Report \cite{LandauerHallIBM55},
Landauer and two collaborators derived some
basic results for the macroscopic Hall effect and electrical conductivity
of an inhomogeneous system constructed of  a conducting material that
has a uniform local Hall  resistivity   $\rho_H^{(M)}$ and a uniform
Ohmic resistivity $\rho^{(M)}$, except for the presence of
non-conducting pores.
(As shown in Ref.\ \cite{BergStrelPRB99_2}, it is not even necessary for $\rho^{(M)}$
to be a scalar quantity.)
In particular, Landauer {\em et al.}  obtained  exact results for
the case where
the microstructure has cylindrical symmetry, i.e., when the microstructure
depends on only two out of three Cartesian coordinates: If the magnetic
field lies along the axis of cylindrical symmetry, then the macroscopic
Hall resistivity $\rho_H^{({\rm eff})}$ is the same as the Hall
resistivity of
the conducting constituent $\rho_H^{(M)}$. If the magnetic field is
perpendicular to that axis, then
\begin{equation}
\rho_H^{({\rm eff})}=\frac{\rho_H^{(M)}}{1-p_M},
\label{HallResistivty}
\end{equation}
where $p_M$ is the volume fraction of the conducting constituent.
These early exact results were later extended
by other groups who studied magneto-transport near a percolation threshold
\cite{StraleyBetheLattice}--\cite{StroudBergmanPRB84}. Other
studies involved
discrete network models which enabled the critical exponent for the weak
(magnetic) field
Hall effect of a percolating system to be evaluated with acceptable
precision
\cite{BergmanKantorStroudWebPRL83,BergmanDueringMuratJStatPhys90},
and scaling theories that discussed connections among different aspects of
the critical behavior \cite{BergmanStroudPRB85,BergmanPhilMag87}.
By extending SEMA to the case where the local resistivity is no longer a
scalar quantity, new critical points were found, as already described
above.
The case of strong (magnetic) field magneto-transport has turned out to
exhibit surprising new features. While the exact results found by Landauer
et al.\ in Refs.\ \cite{LandauerHallJAP56,LandauerHallIBM55} are
valid for arbitrary field strength, other results have been found more
recently which are asymptotically exact only in the strong field limit,
i.e., when the Hall-to-Ohmic resistivity ratio in at least one constituent
is much greater than 1 \cite{BergStrelPRL98_1}--\cite{BergmanCrete}.

In 1957, Landauer wrote a pioneering paper on electronic transport
due to localized scatterers in a metal. In this article he pointed
out that,
besides the scattering of individual electrons by the impurity potential,
another important effect must also be taken into consideration, namely,
the inhomogeneity of electron density induced by that potential.
That effect had previously been ignored. This article, which appeared in
the first volume of the IBM Journal of Research and Development
\cite{LandauerMigration57}, was not properly appreciated
until it transpired that the insights developed in it are
extremely relevant for understanding the phenomenon of
electromigration---see Ref.\
\cite{SorbelloReview97}
for a detailed list of relevant references on this topic.
Because the availability of Ref.\ \cite{LandauerMigration57} was
so limited, the unusual step was taken of re-publishing it as an
article in the Journal of Mathematical Physics nearly 40 years
after the original publication---see Ref.\ \cite{LandauerJMathPhys96}.
This is just one example of how far ahead of most other physicists
Landauer was
in his scientific thinking and insight: Until others caught up with
his 1957 results, the original work had almost vanished into
oblivion.

The phenomenon of electromigration was actually a major interest of
Landauer for much of his life.
The subject was of great practical importance to IBM, as a principal
mechanism for the  failure of integrated circuits is deterioration caused
by electromigration of  defects and impurities near junctions in the
circuit. Landauer's focus was on the microscopic understanding of forces
responsible for the motion of defects.  An important early contribution
to this field was the paper by Landauer and Woo, ``Driving force in
electromigration'', published in 1974 \cite{L47}.  The central idea of this
paper was that the inhomogeneity in the electron density near a defect
or impurity carries with it a change in the local conductivity. When an
electric current is applied, this leads to formation of electric dipoles,
which can exert a force on the defect, in addition to forces resulting
from the direct transfer of momentum from an electron to the impurity
during a scattering process.  The issue confronting Landauer and Woo was
how to properly take this force into account.  Landauer wrote a number of
subsequent papers on the driving forces for electromigration, which we
will not have room to summarize here. However, the interested reader can
find a review of  Landauer's contributions to the subject in an article by
R. Sorbello,   entitled ``Landauer fields in electron transport and
electromigration", published in 1998 \cite{Sorbello98}.

When the first ETOPIM conference was convened, in Columbus, OH during
7--9 September 1977 \cite{ETOPIM1}, Rolf Landauer was asked to deliver
the opening keynote address. In that talk, he presented an exhaustive
review of the development of theoretical treatments for the physical
properties of a composite medium up to that time. The article which
summarizes that talk in the conference proceedings volume
\cite{LandauerETOPIM1} is an invaluable historical review, which also
lists
and discusses all the important articles in that field which were known at
that time---altogether 163 references.

\section{Quantum systems}
\label{Quantum Systems}

At an early stage of his career, Landauer became interested in the study
of  systems of non-interacting electrons in a  one-dimensional disordered
potential, which could be studied analytically or numerically with the
computers of the time, and could  shed light on more realistic
three-dimensional systems which were then not tractable.
The work of Landauer
and Helland, in  1953, was a pioneering work in this area \cite{L10}.
However, the most influential paper that Landauer wrote based on the
analysis of one-dimensional systems was his 1970 paper,
``Electrical resistance of disordered one-dimensional lattices''
\cite{LandauerFormulaPhilMag69}.
 The 1970 paper was important  because of its contribution to our
 understanding of the phenomenon of localization in one-dimensional
 systems, but even more significantly, it established a connection
 between electrical conductance and transmission probabilities, that
 has been  the basis for much future work on mesoscopic systems, often
 referred to as the Landauer formalism.

What Landauer did in the 1970 paper was to  study statistical properties
of the transmission matrix through a one-dimensional region with a sequence
of partially reflecting barriers randomly spaced. As Landauer noted, if
the disordered region (let us call it the ``sample'')  is connected on
either end by smooth wires to reservoirs at different chemical potentials,
there will be a net current through the sample determined by the potential
difference of the reservoirs and the  transmission probability  $T$
for an electron with energy close to the Fermi energy, incident on
the sample from either side.  (It is a consequence of the principle
of detailed balance that the transmission probability will be the same
whether the  electron is incident from left or right.)  Landauer used
this result to define a conductance for the system, which he found to be
\begin{equation}
G = \frac{e^2}{h} \, \, \frac {T} {1-T} .
\label{eqG}
\end{equation}
The formula is for spinless non-interacting electrons, in the limit of
zero temperature.
The transmission probability is, in turn, related to the complex
transmission amplitude $t$, by $T=|t|^2$.

A decade after Landauer's formulation, an alternative relation
between conductance and transmission probability was proposed by
Economou and Soukoulis \cite{EcSouk}  and others. (See the discussion
in Ref. \cite{ImryBook}, particularly pages 93--103.)   For spinless
electrons in one dimension, this relation is simply
\begin{equation}
\Gamma = \frac{e^2}{h} T
\label{eqGamma}
\end{equation}
This formula was also generalized to the case where there can be several
transverse channels for electrons in the wires connected to the sample. In
this case we have \cite{FisherLee}
\begin{equation}
\Gamma= \frac{e^2}{h} \sum_{ij} |t_{ij}|^2 \, ,
\label{multichannel}
\end{equation}
where $i$ and $j$ label the channels in the left and right leads
respectively, and $t_{ij} $ is the matrix of  transmission amplitudes.

For some time, there was much discussion about which of the two
quantities, $G$ or $\Gamma$, is the ``correct'' definition of the
electrical conductance.  We now understand that they are, in some sense,
both correct, but refer to different experiments \cite{ImryBook, L130}.
The quantity $\Gamma$ should be thought of as a two-terminal conductance.
If the sample is connected by ideal wires to two large reservoirs,
in equilibrium at voltages $V_1$ and $V_2$,  and $I$ is the current
through the sample, then
\begin{equation}
\Gamma = \frac{I}{V_1 - V_2} \, .
\end{equation}
In contrast,  $G$ may be thought of as a four terminal conductance.
If one could attach an ideal voltage probe to the leads on  either side
of the sample, which would measure the voltages  $V_3$ and $V_4$ in the
leads without drawing any current from the leads and without disturbing
them in any way, we would have
\begin{equation}
G = \frac{I}{V_3 - V_4} \, .
\end{equation}

Unfortunately, it  is not entirely clear how one could construct an ideal
voltage probe that would not disturb a mesoscopic system \cite{ImryBook}.
In fact, since the electrons within the leads are not in  thermal
equilibrium, there may be some question how one should properly define
a voltage in the leads.  Landauer had in mind that the voltage would
be defined by the total density of left and right moving electrons,
as well as by the electrostatic potential, which should be determined
self-consistently.  On the other hand,  it has proved relatively easy to
fabricate mesoscopic systems with good connections to external reservoirs
of known voltage, so the two-terminal conductance $\Gamma$ has proved
to be an extremely useful concept.
Despite the difference between $\Gamma$ and the conductance $G$ that
Landauer originally introduced, Landauer deserves a great deal of credit
for introducing the idea that the conductance should be determined by
the transmission probabilities.

In Landauer's original paper, and much of the subsequent work, analysis
was restricted to non-interacting electrons, or models where the
Coulomb interaction is  introduced only in the form of a self-consistent
potential.  However, we understand that the analysis is also applicable
for interacting electron systems, provided that the temperature is low
and the system sufficiently small so that electrons that enter the sample
will leave it before suffering an inelastic collision. Since the time for
inelastic collisions increases as the temperature is  reduced, studies
of these phenomena are generally carried out at very low temperature.

Landauer's 1970 paper had importance separate from the general question
of conductance through mesoscopic systems.  The paper shed very important
light on the issue of electron localization in one-dimensional systems.
In previous work, by Mott and Twose, by Borland, and by others, it had
been established that for non-interacting electrons in a disordered
potential in one dimension, in the limit of an infinite wire, the
electron eigenstates would all be localized, except for a possible set
of measure zero \cite{Mott, Borland, Halp}. What this meant was that for
each eigenstate, there would be a point on the line where the magnitude
had a maximum, and on either side of that point, the wave function
would decrease exponentially, with a decay length that depended on the
energy and the strength of the disorder, but would remain  finite in the
limit of an infinite system.  As a result of this, it was argued that the
resistance of a long one-dimensional disordered system of non-interacting
electrons should  increase exponentially with the length $L$, in contrast
to a classical wire, where the resistance is linear in $L$.  Landauer
was able to explain the exponentially diverging resistance in terms of
transmission amplitudes and quantum mechanical interference in the wire.

To understand Landauer's argument, let us consider his formula, Eq. (5) of Ref.~\cite{LandauerFormulaPhilMag69}, 
for the
inverse of the  conductance $G$ of a sample consisting of two barriers
in series:
\begin{eqnarray}
\lefteqn{\frac{e^2}{hG} = \frac{1-T}{T} =}\nonumber\\
&& \hspace{-13 true mm}\frac{ (1-T_1) +(1-T_2) + 2
(1-T_1)^{1/2}   (1-T_2)^{1/2} \cos \phi     }{T_1 T_2}.\nonumber\\
&&\label{twobarriers}
\end{eqnarray}
Here $T$ is the transmission of the system as a whole, $T_1$ and $T_2$
are the transmissions of the individual  barriers, and the phase $\phi$
depends on the distance between barriers.  That phase arises from
the interference of contributions in which the particle is reflected
multiple times by the barriers before finally emerging from one side
or the other of the system.  Formulas for three or more barriers can be
obtained by iteration, adding one barrier at a time.

To introduce the effects of disorder, Landauer  considered a model of $N$
barriers having  identical individual transmission  probabilities  $T_1$,
but with random spacings among them.  In particular, Landauer assumed
that the phases $\phi$ between successive barriers could be treated as
independent random variables, uniformly distributed from 0 to $2 \pi$.
This can be strictly justified when the variation in the distance between
barriers is large on the scale of the wavelength of the electrons,
but the final results are actually much more general. Landauer showed
that the mean value of the  resistance $1/G$ for his model   is  given
by the formula
\begin{equation}
\left\langle \frac{e^2}{hG} \right\rangle =
\frac{1}{2} \left[\left( \frac{2-T_1}{T_1 } \right)^N -1 \right]\;\;.
\label{Nbarriers}
\end{equation}
It follows that the mean value of the resistance will diverge
exponentially with the length of the system, unless $T_1=1$, {\em i.e.},
unless there is perfect  transmission for the individual  barriers.

When the resistance of the sample is very large, it does not matter
whether one  considers the four-terminal resistance $1/G$ or the
two-terminal resistance $1 / \Gamma$.  The mean values of both  quantities
will diverge, at the same exponential rate,  as $N$ becomes large. By
contrast, the mean values of $G$ and $\Gamma$ are quite different. Though
typical values of $G$ will be exponentially small, as expected from
the large value of $\langle G^{-1} \rangle$,  the mean value of $G$,
for a sample of specified length $N$,  will actually be infinite, as
Landauer noted in his paper:
\begin{equation}
\langle G \rangle = \left\langle \frac{T}{1-T} \right\rangle = \infty \, .
\end{equation}
The reason for this can be seen by inspection of
Eq. (\ref{twobarriers}). The formula implies that when $\Delta T \equiv T_1-T_2$ and 
$\Delta \phi \equiv \phi - \pi$ go to zero, 
the value of $G$ will diverge as $[(\Delta T)^2 + (\Delta \phi)^2]^{-1} $.  Let us divide our sequence of $N$ barriers into two roughly equal halves, let $T_1$ and $T_2$ be the separate transmission  probabilities of the two halves, and let $\phi$ be the phase accumulation in the space between the two halves.  As the probability density for $\Delta T$ and $ \Delta \phi$ will, in general, be finite when the two variables go to zero,  the mean value of $G$ will diverge logarithmically.

\subsection{Applications}
\begin{figure}
\includegraphics[width=\columnwidth]{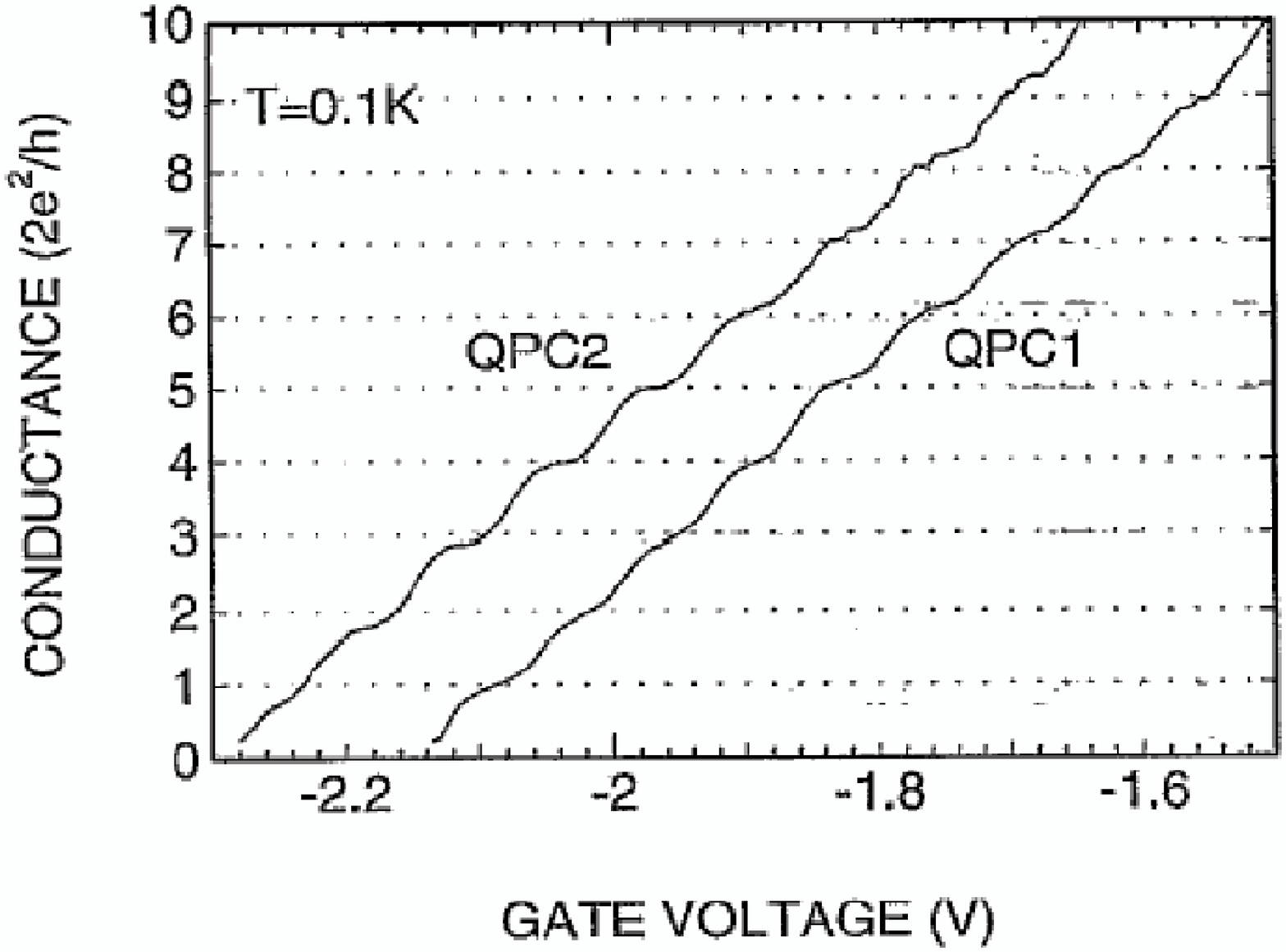}

\vspace{10 true mm}

\hspace{14 true mm}\includegraphics[width=6 true cm]{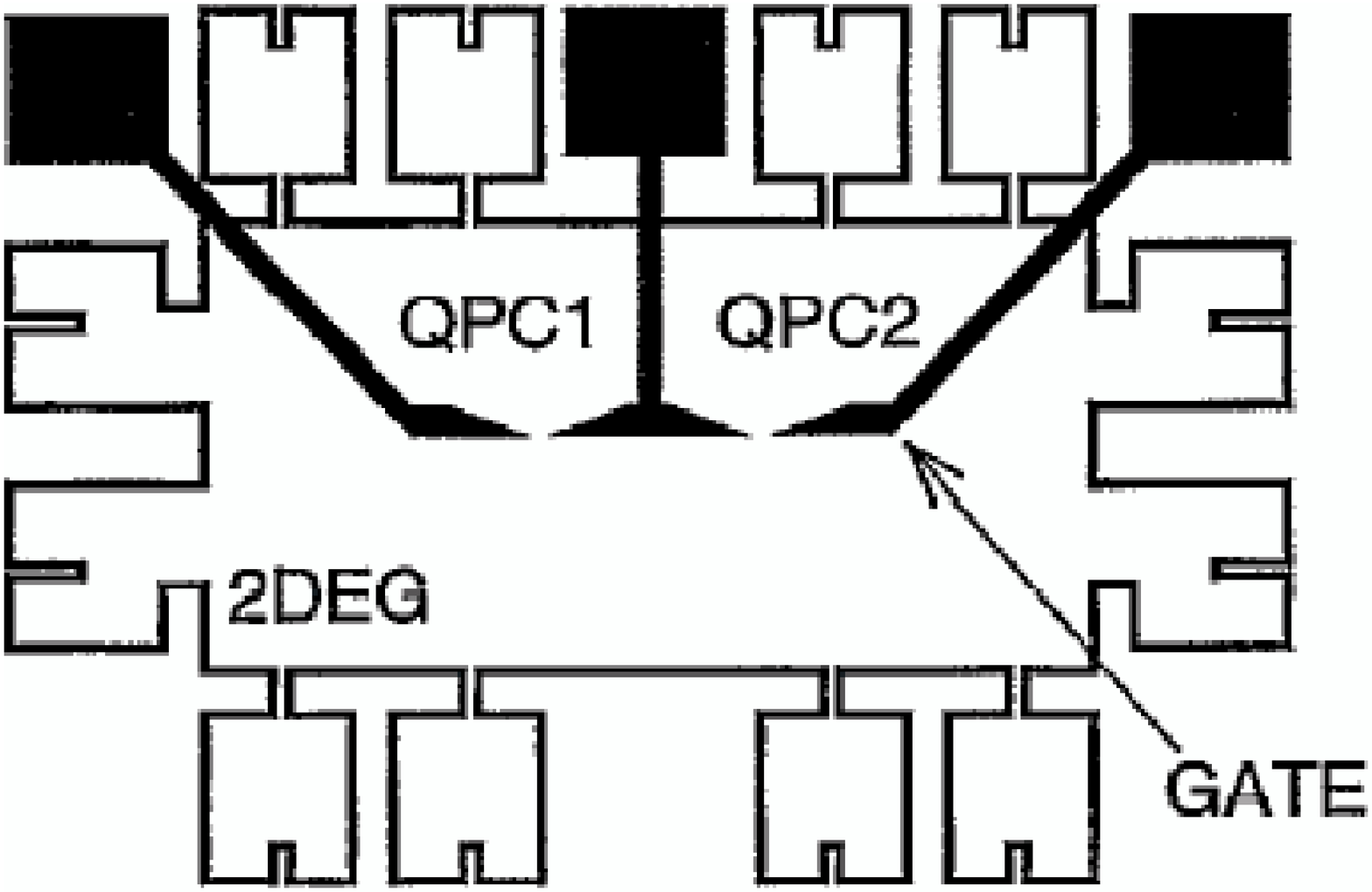}
\caption{
Conductance steps in a two-dimensional electron system with a
constriction of variable width. Upper panel shows measured
conductances  of two constrictions, QPC1 and QPC2, versus the
voltage applied to gate electrodes that control the widths of
the constrictions.  Lower panel is a schematic of the experimental
system.  Reprinted from Ref.\ \protect\cite{Gao} with permission from
The American Institute of Physics.
\label{ConductivitySteps}}
\end{figure}

An important application of the Landauer-type approach to conductance
resulted from the dramatic experimental discovery in 1988 of
quantized conductance steps in  semiconductor devices with a narrow
constriction \cite{Wees, Wharam, Gao}.  In these devices, fabricated from
two-dimensional electron systems in GaAs,  the width of the constriction
could be varied continuously by applying a negative bias to a pair of gate
electrodes on the surface of the sample---see lower panel of
Fig.\ \ref{ConductivitySteps}. As the width of the constriction
was varied, the conductance was not a linear function of the gate voltage,
but was seen to exhibit a series of plateaus, with values
$
\Gamma = 2 N e^2/h \, ,
$
where $N$ is a positive integer---see upper panel of
Fig.\ \ref{ConductivitySteps}.
These observations could be understood
using Landauer's ideas, if we assume that for electrons in a given channel
of transverse motion, as soon as the constriction is wide enough to
permit transmission of  electrons at the Fermi energy, the transmission
probability $T$ is very close to unity.  If the constriction is too
narrow, then transmission in the given channel will be close to zero.
Thus, the conductance will be close to an integer times $e^2/h$.
The factor 2 appears because of the degeneracy due to electron
spin.  The sudden increase in $T$ from complete  reflection to complete
transmission is very plausible in these systems because the controlling
gates are set back from the two-dimensional electron gas by a distance
large compared to the Fermi wave length.  Thus the potential felt by the
electrons should be very smooth, and the the transmission problem reduces
to the semiclassical problem of a particle incident upon a barrier, where
the transmission probability is either 0 or 1, depending on whether the
particle has enough energy to get over the barrier.  In the years since
1988, an enormous number of experimental and theoretical investigations
have been built on these experiments and their interpretation.




Landauer's ideas were also important in the understanding of the
resistance oscillations of a microscopic metal ring attached to two
leads, as a function of the magnetic flux threading the ring.  The key
theoretical paper here was the work of B\"utikker, Imry, Landauer and
Pinhas, in 1985 \cite{L101}.
This work was, in turn, closely tied to  experiments carried out at
IBM at that time \cite{Webb85}, which studied the magnetoresistance of
a thin gold ring, approximately 800 nm in diameter, with a thickness of
approximately  40 nm. Upon varying the magnetic field, the experiments
found oscillations in the  resistance corresponding to a fundamental
period of the addition of one quantum of magnetic flux, $\Phi_0 = h/e$,
through the hole in the ring. Fourier transform of the data showed strong
peaks at  frequencies corresponding to this fundamental, and also at
the first harmonic, corresponding to addition of half a flux quantum.
As explained by B\"uttiker et al.\ \cite{L101},
the fundamental frequency in the
transmission amplitude arises from the quantum interference between
paths in which an electron may travel from one contact to the other along
either side of the ring.  The phase of this interference term depends on
the precise location of scattering centers in the sample, and would be
expected to vary randomly from one
sample to another.  Thus oscillations at the fundamental period would be
expected to vanish, or be greatly reduced, in an experiment where the
signal was averaged over many different rings,  However, oscillations
corresponding to one-half flux quantum would not vanish on averaging,
and thus would dominate an averaged measurement.  This analysis was
consistent with the results of previous experiments and on multiple
rings, and on tubular samples, which may be thought of as many rings in
parallel \cite{SharvinSharvin, Altshuler, Pannetier}.

Another geometry that interested Landauer was the case of an isolated
metal ring, with no electrical contacts.  Classically, the conductance
of a closed metal ring can be measured by placing it in a time-varying
external magnetic field and measuring the  magnetic moment induced in
the ring.  The induced moment would be proportional to the current
flowing around the ring, which in turn would be proportional to the
time-derivative of the flux and the conductance of the ring.  For a
mesoscopic wire at low temperatures, where the discrete quantization of
electronic levels becomes important,  the situation is more complicated.
In this case, there can be a non-zero ``persistent current", in
equilibrium in a dc magnetic field, which will be an oscillatory function
of  the flux through the loop. Such persistent currents are well known
in superconductors, but they also occur (with much smaller magnitudes)
in normal metal loops.  Landauer and his collaborators wrote a number
of important papers in the 1980s which discussed both the existence and
magnitude of persistent currents in a dc magnetic field, and the
behavior to be expected in a time-varying magnetic field
\cite{L92, L100, L108, L113, L114}.
Landauer's analysis of the latter problem also allowed him to address
fundamental issues of the nature of dissipation in small closed loops.

Landauer's approach to conductance was also the basis for important
work on shot noise in mesoscopic systems.  The formula for the current
noise-power, per unit frequency, in the limit of zero temperature,
is given by \cite{Khlus, Lesovik, ButNoise, Yurke}
\begin{equation}
S = 2 (e^2/h) V \sum_i T_i (1-T_i) \, ,
\label{noise}
\end{equation}
where $V$ is the applied voltage and $T_i$ is the transmission probability
for the $i$th channel. [Here, we have made a unitary transformation on
the channels in the two leads so that the transmission matrix  $t_{ij}$
is diagonal, and the conductance formula (\ref{multichannel}) becomes
$\Gamma =
(e^2/h) \sum_i T_i$.]  The noise formula (\ref{noise}) is very widely
used, and has been the basis for much subsequent work.  Landauer's own
views on  shot noise may be found in his article ``Mesoscopic Noise:
Common Sense View'', published in 1996 \cite{L171}. (See also the 1991
article by Landauer and Martin \cite{L146}.)

Another interest of Landauer, related to mesoscopic systems, was the
concept  of {\em transit time} in tunneling events.  The reader is
referred to Refs. \cite{L150, L153} for  Landauer's views on this subject.

\section{Biographical summary}

Rolf Landauer was born in Stuttgart, Germany, in 1927.  He moved to
the United States, with his family, in 1938, several years after the
death of his father in 1935.   His father, who had fought for Germany
in World War I, and had been severely wounded, was very patriotic, and
did not want to leave the country.  He strongly  believed that the Nazi
antisemitism would pass.
Landauer has said that were it not for the early death of his father,
due in part to problems resulting from  his war wounds,  his family
would have undoubtedly remained in Germany until it was too late to leave.

Landauer's family settled in New York City, where he went to high school,
before entering Harvard College at  the age of 16.  He graduated in two
years, and enlisted in the U. S. Navy, where he claimed to have learned as
much as he had learned in college.  Eventually, he returned to Harvard for
graduate studies, where he received his Ph.D. in 1950.  After graduate
school,  Landauer worked for two years at the Lewis Laboratory of the
National Advisory Committee for Aeronautics (NACA, later to be renamed
NASA, acronym for the National Aeronautics and Space Administration).
In 1952, he moved  to IBM Laboratories in Poughkeepsie, NY, (later to be
renamed the IBM Thomas J. Watson Research Center),  where he continued
to work until he passed away in 1999.

At IBM, in addition to his research work, Landauer held important
management posts at various times.  He was responsible for much of IBM's
early work on large scale integration, and has been given credit for
inventing the term.
Landauer was awarded the title of ``IBM Fellow'' in 1969.

\section*{Acknowledgments}

Research of BIH was supported in part by the National Science Foundation  (grant DMR-09-06475).  Research of DJB was supported by the Israel Science Foundation
(grant No.\ 585/06).



\end{document}